\documentclass{article}

    \PassOptionsToPackage{square,sort,comma,numbers}{natbib}

    \usepackage[final]{neurips_mlsb_2020}

\usepackage[utf8]{inputenc} 
\usepackage[T1]{fontenc}    
\usepackage{hyperref}       
\usepackage{url}            
\usepackage{booktabs}       
\usepackage{amsfonts}       
\usepackage{nicefrac}       
\usepackage{microtype}      
\usepackage{multirow}
\usepackage[pdftex]{graphicx}
\usepackage[frozencache,cachedir=.]{minted}
\usepackage{svg}
\usepackage{subcaption}
\usepackage{verbatim}

\title{SidechainNet: An All-Atom Protein Structure Dataset for Machine Learning}

\author{%
  Jonathan E. King\thanks{Carnegie Mellon University-University of Pittsburgh Joint PhD Program in Computational Biology} \\
  Comp. \& Systems Biology \\
  University of Pittsburgh \\
  Pittsburgh, PA 15213 \\
  \texttt{jok120@pitt.edu}
  \And
  David Ryan Koes \\
  Comp. \& Systems Biology \\
  University of Pittsburgh \\
  Pittsburgh, PA 15213 \\
  \texttt{dkoes@pitt.edu} \\  
}

\begin{document}

\maketitle
\begin{abstract}
Despite recent advancements in deep learning methods for protein structure prediction and representation, little focus has been directed at the simultaneous inclusion and prediction of protein backbone and sidechain structure information. We present SidechainNet, a new dataset that directly extends the ProteinNet dataset. SidechainNet includes angle and atomic coordinate information capable of describing all heavy atoms of each protein structure. In this paper, we provide background information on the availability of protein structure data and the significance of ProteinNet. Thereafter, we argue for the potentially beneficial inclusion of sidechain information through SidechainNet, describe the process by which we organize SidechainNet, and provide a software package (\texttt{\href{https://github.com/jonathanking/sidechainnet}{https://github.com/jonathanking/sidechainnet}}) for data manipulation and training with machine learning models. 
\end{abstract}

\section{Background}

The deep learning subfield of protein structure prediction and protein science has made considerable progress in the last several years. In addition to the dominance of the AlphaFold deep learning method in the 2018 Critical Assessment of protein Structure Prediction (CASP) competition \cite{AlphaFoldNews, Senior2020, CASP13}, many novel deep learning-based methods have been developed for protein representation \cite{ProteinRep, ingraham2019generative}, property prediction \cite{SureyyaRifaioglu2019, DeepFam}, and structure prediction \cite{Xu2019, Yang2020, AlQuraishi2019, ingraham2018learning}. Such methods are demonstrably effective and extremely promising for future research. They have the potential to make complex inferences about proteins much faster than competing computational methods and at a cost several orders of magnitude lower than experimental methods.

\hypertarget{sectiononeone}{}
\subsection{Protein Structure Data Availability and Information Leakage}
The availability of existing data is a notable challenge for applications of deep learning methods to protein science and, in particular, to protein structure prediction. This limitation is not equally present in other applications of deep learning. For instance, in the field of image recognition, the ImageNet \cite{imagenet} dataset contains 14 million annotated and uniformly presented images. Similarly, in the field of natural language processing, linguistic databases and the web provide access to hundreds of millions of samples of annotated textual data and effectively limitless access to unannotated data. Though the Universal Protein Resource (UniProt) database now contains a whopping 156 million unique protein sequences \cite{uniprot}, there are currently only 170,968 protein structures in the Protein Data Bank \cite{ThePDB} (PDB), of which only 98,347 represent unique protein sequences. Furthermore, although protein structure data is accessible through the PDB in individual files, it is not otherwise preprocessed for machine learning tasks in a manner analogous to other machine learning domains.

In addition to data availability, biased or skewed data presents another major challenge for machine learning practitioners \cite{standards}. Something as simple as dividing the data into training and testing splits for model development and evaluation can introduce harmful biases. Failure to analyze the similarities or differences between training and evaluation splits can lead to “information leakage” in which trained models exhibit overly optimistic performance by learning non-generalizable features from the training set. One such example of this in structural biology research is the Database of Useful Decoys: Enhanced (DUD-E) dataset \cite{dude}. DUD-E remains a common benchmark for evaluating molecular docking programs and chemoinformatic-focused machine learning methods despite recent research \cite{dudeissues,sieg2019need} uncovering dataset bias that explains misleading performance by many of the deep learning methods trained on it.

\hypertarget{sectiononetwo}{}
\subsection{Treatment of Protein Backbone and Sidechain Information}
A common way to describe the structure of a protein is to divide it into two separate components–the backbone and the sidechains that extend from it. The protein backbone is a linear chain of nitrogen, carbon, and oxygen atoms. The torsional angles ($\Phi, \Psi$, and $\Omega$) that connect these atoms form the overall shape of the protein. In contrast, protein sidechains are chemical groups of zero to ten heavy atoms connected to the central $\alpha$-carbon of each amino acid residue. Each of the twenty distinct amino acids is defined by the unique structure and chemical composition of its sidechain component. Consequently, each protein is defined by the unique sequence of its constituent amino acids. The precise orientation of amino acid sidechains is critical to the biochemical function of proteins. Enzyme catalysis, drug binding, and protein-protein interactions all depend on a level of atomic precision that is not accounted for in backbone structure alone. Thus, the protein backbone and sidechain are both crucially important to protein structure and function.

Despite this, many predictive methods treat backbone structure prediction and sidechain structure prediction as distinct problems.\footnote{Recent work by Yang et al. \cite{Yang2020} included a representation of inter-residue $\beta$-carbon orientations, but did not provide a representation of complete sidechain structures.}  For instance, a common approach is to predict the protein backbone alone and then add sidechains to the generated structure through a conformational or energy-minimizing search \cite{scpacking}. Other approaches, through tools such as Rosetta \cite{rosettaCASP8}, optimize the structure using backbone dependent rotamer libraries \cite{Shapovalov2011}. In both cases, sidechain placement is dependent on predicted backbone structure.

The methods currently employed by the community are undoubtedly effective. A possible reason for separating backbone and sidechain information could be that the community believes adequate progress in backbone structure prediction has been made without accounting for sidechain conformations during training. Alternatively, it may be the case that such information is simply not readily available for training deep learning models. Nonetheless, the potentially positive effect of including the complete backbone and sidechain structure at training time has not been determined.

\section{SidechainNet}
We propose SidechainNet, a protein sequence and structure dataset that addresses the concerns described in Sections \hyperlink{sectiononeone}{1.1} and \hyperlink{sectiononetwo}{1.2}.

\subsection{Addressing Protein Structure Data Availability and Information Leakage}
SidechainNet is based on Mohammed AlQuraishi's ProteinNet  \cite{proteinnet}. ProteinNet is a dataset designed to mimic the assessment methodology of CASP proceedings. 

Under the CASP contest organization, participants develop predictive methods using any publicly available protein structures. Thereafter, predictive methods are assessed using a specific set of proteins that are selected by contest organizers and withheld from participants until assessment. Organizers select proteins that are challenging to predict and, in the case of the Free Modeling contest category, have minimal similarity to available protein structures. In effect, the CASP process results in a portion of data for method development and a separate portion for evaluation. AlQuraishi proposed that these subsets could be treated as training and testing sets and re-purposed for machine learning.

Since no validation sets are constructed by the CASP organizers, AlQuraishi used clustering methods to extract distinct protein sequences and structures from each training set to use for validation. For each CASP contest, AlQuraishi ultimately constructed seven different validation sets–each containing an increasing upper bound on the similarity between sequences in the given validation set and those in the training set. His construction allows users to evaluate a model’s performance on proteins that have high similarity to a given training set (akin to the Template-Based Modeling CASP category) or proteins that have low or zero similarity to a training set (akin to the Free Modeling CASP category). 

 As a result of the constraints imposed by the CASP contest structure and AlQuraishi's validation set construction, ProteinNet accounts for protein sequence similarity across data splits. This prevents information leakage and minimizes misleading performance. For these reasons, we have replicated AlQuraishi's training, testing, and validation set constructions in SidechainNet.
 
\subsection{Unifying Protein Backbone and Sidechain Information}

SidechainNet also extends ProteinNet by including all heavy atoms of each protein and all necessary torsional angles for atomic reconstruction. 

We are interested in determining whether the availability of complete sidechain conformation information can improve the performance of structure prediction methods. However, even if such models are not more accurate, they will be fundamentally more expressive as they will generate all-atom protein structures. These structures could be used without modification in downstream analyses for structure-based drug discovery and even molecular dynamics simulations.

\begin{table}[b]
\centering
\begin{tabular}{cccc}
\textbf{Entry}                       & \textbf{Dimensionality} & \textbf{ProteinNet} & \textbf{SidechainNet} \\ \hline
Primary sequence                     & $L\times 1$             & X                   & X                     \\
PSSM$^\dagger$ + Information Content & $L\times 21$            & X                   & X                     \\
Missing residue mask                 & $L\times 1$             & X                   & X                     \\
Backbone coordinates$^*$             & $L\times 4\times 3$     & X                   & X                     \\ \hline
Backbone torsion angles              & $L\times 3$             &                     & X                     \\
Backbone bond angles                 & $L\times 3$             &                     & X                     \\
Sidechain torsion angles             & $L\times 6$             &                     & X                     \\
Sidechain coordinates                & $L\times 10\times 3$    &                     & X                    
\end{tabular}
\caption{\label{tab:table-name}Differences between ProteinNet and SidechainNet where $L$ represents protein length. $^\dagger$Position Specific Scoring Matrices (PSSMs) developed from multiple sequence alignments. \newline$^*$SidechainNet includes atomic coordinates for backbone oxygen atoms while ProteinNet does not.}
\end{table}

\begin{figure}
	\centering
    \includegraphics[width=0.99\linewidth]{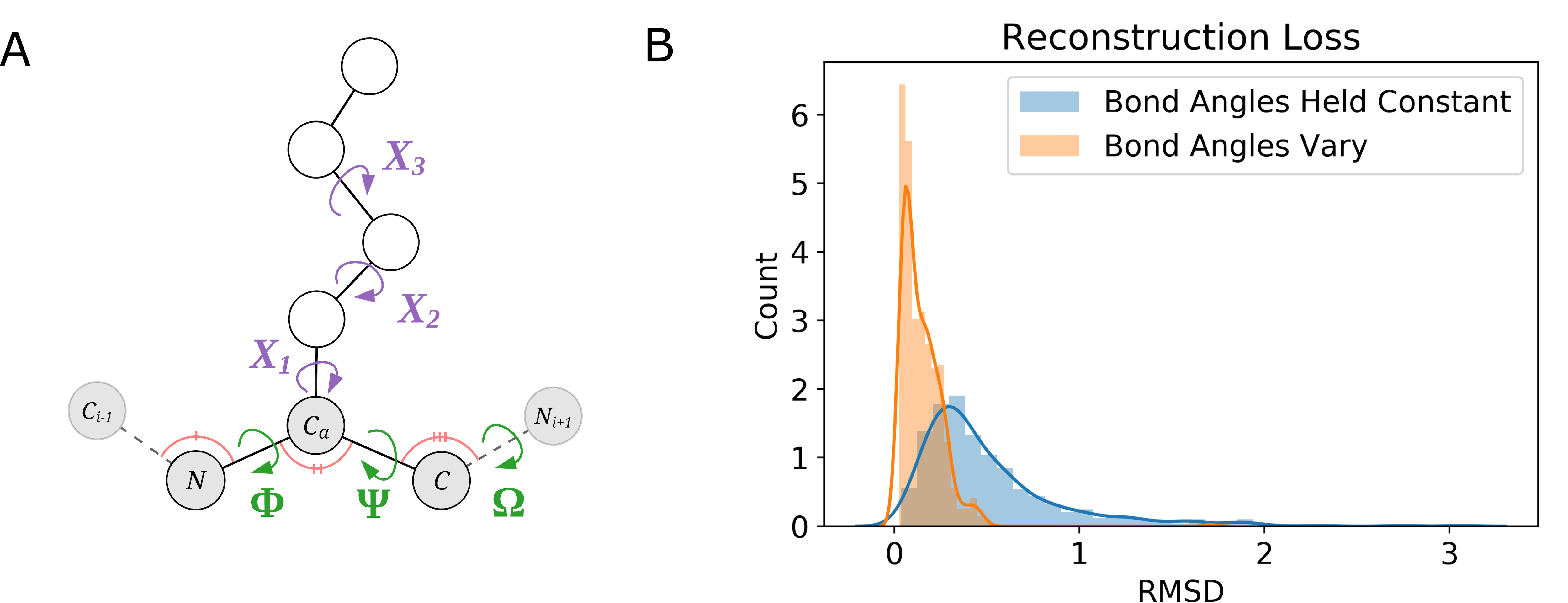}
	\caption{ \textbf{(A)} A methionine residue annotated with the 3 categories of angles measured by SidechainNet: the canonical backbone torsional angles ($\Phi, \Psi$, and $\Omega$) in green, the rotatable sidechain-specific torsional angles ($X_{1-3}$) in purple, and the 3-atom backbone bond angles in pink. Atoms considered part of the residue's sidechain are colored white, while the backbone atoms are colored gray. \newline\textbf{(B)} Backbone bond angles (\texttt{C-N-CA}, \texttt{N-CA-C}, and \texttt{CA-C-N}; colored pink in \ref{fig:ncac}\textbf{A}) are important for accurately reconstructing atomic coordinates from angles. When we assumed that these angles could be fixed during coordinate reconstruction, the result was greater reconstruction error (blue distribution). When we measured these angles with SidechainNet and included their actual values during coordinate reconstruction, we achieved lower reconstruction error (orange distribution). Root-Mean-Square Deviation (RMSD) measures the difference between true atomic coordinates and coordinates reconstructed from recorded angles.}
	\label{fig:ncac}
\end{figure}

\section{Properties of SidechainNet}
To the maximum extent possible, SidechainNet attempts to replicate the features included in ProteinNet (Table \ref{tab:table-name}). However, SidechainNet adds support for several new features as well. More information about the new features is included below.

\subsection{Angle Information}

One important component of protein structure data often utilized in structure prediction methods but absent from ProteinNet is the set of torsional angles that describe the orientation of the protein. The canonical backbone torsional angles ($\Phi, \Psi$, and $\Omega$) for each residue are provided in SidechainNet.

In the development of SidechainNet, we were surprised to observe that we incurred a noticeable amount of error when attempting to reconstruct a protein's Cartesian coordinates given only its backbone torsional angles (Figure \ref{fig:ncac}B). We determined that the error was due to our assumption that several important non-torsional angles in the protein backbone (\texttt{C-N-CA}, \texttt{N-CA-C}, and \texttt{CA-C-N}) could be fixed using reference values from AMBER force fields \cite{amber} rather than being allowed to vary. Thus, even if a model perfectly predicted backbone torsional angles, it would still be subject to reconstruction error if such bond angles were held fixed. For that reason, we decided to include these angles in SidechainNet as well (Figure \ref{fig:ncac}A).

\subsection{Sidechain Information}
SidechainNet also includes sidechain-specific information that describes the orientation of each amino acid using both external (Cartesian) and internal (angular) coordinate systems. This information includes up to ten atomic coordinates and up to six torsional angles for the sidechain component of each residue.  

The number of angles measured for each amino acid sidechain depends on the number of rotatable bonds in the amino acid. For those residues with aromatic sidechains (e.g., tryptophan), the torsional angles describing the ring components of the residue are omitted since they can be inferred, but their complete atomic coordinates are still recorded. 

3-atom bond angles (as opposed to 4-atom torsional angles) and bond lengths associated with sidechain structures are not recorded from structure files. Instead, we extracted reference values unique to each bond from the AMBER \texttt{ff19SB} force field \cite{amber, amberff19sb}. Together with the recorded sidechain torsional angles, these angles and bond lengths can be used to generate all-atom sidechain structures.

\subsection{Generation Procedure and Caveats}

We constructed SidechainNet by, first, obtaining raw ProteinNet text records linked to in ProteinNet's GitHub repository (\href{https://github.com/aqlaboratory/proteinnet}{https://github.com/aqlaboratory/proteinnet}). Then, we parsed these records before saving them into Python dictionaries. Next, we re-downloaded all-atom protein structures (sequences, coordinates, and angles) for each protein described by ProteinNet from the PDB using the ProDy software package \cite{Bakan2011}. Finally, we combined the remaining data (PSSMs and Information Content values) into the final SidechainNet dataset by aligning the sequences observed during the re-downloading process with the sequences described by ProteinNet. We replaced the coordinate and missing residue information described in ProteinNet  with the re-downloaded values to ensure consistency. We also temporarily excluded secondary structure information, which is absent from the ProteinNet release as of writing. Scripts and instructions for constructing SidechainNet data are available in our GitHub repository.

We failed to include data for 326 (0.3\%) of the 104,323 listed protein structures from ProteinNet's CASP12 dataset. Issues with these structures arose due to problematic file parsing, edge cases, and disagreement between the sequences we observed and those reported by ProteinNet that could not be resolved programmatically. Although we currently exclude these protein structures, we plan to resolve remaining issues in a subsequent release.


Furthermore, both ProteinNet and SidechainNet only consider single-molecule protein chains. Proteins from the PDB that contain multiple chains are divided into multiple independent entries, one for each chain. Any missing residue or atomic information is padded with zero-vectors matching the size of the corresponding data (e.g., $1\times3$ for each missing atom, $1\times1$ for each missing angle). In addition, detailed characteristics of structure data (e.g. b-factors, multiple or alternative sidechain locations, and resolution) are ignored.

Lastly, structures are recorded directly from the PDB and have no guarantee to be energetically minimized. In a future release, we plan to use AMBER to energetically minimize the structures, as this may make the data more consistent and amenable to training. It will also improve our ability to accurately reconstruct protein structure coordinates from angles.

\section{Using SidechainNet}

SidechainNet was originally developed for Python and the PyTorch machine learning framework. As such, we have developed a Python package to load SidechainNet and efficiently train models with it.

\hypertarget{fourone}{}
\subsection{Loading SidechainNet as a Python dictionary}

In its simplest form, SidechainNet is stored as a Python dictionary organized by the same training, validation, and testing splits described in ProteinNet. There are multiple validation sets included therein.
 
 Within each of SidechainNet's training, validation, and testing splits is another dictionary mapping data entry types (\texttt{seq}, \texttt{ang}, etc.) to a list containing this data type for every protein. In the example below, \texttt{seq\{i\}}, \texttt{ang\{i\}}, ... all refer to the \texttt{i}$^{th}$ protein in the dataset.

\begin{minted}{python}
data = {"train": {"seq": [seq1, seq2, ...],  # Sequences
                  "ang": [ang1, ang2, ...],  # Angles
                  "crd": [crd1, crd2, ...],  # Coordinates
                  "msk": [msk1, msk2, ...],  # Missing residue masks
                  "evo": [evo1, evo2, ...],  # PSSMs and Information Content
                  "ids": [id1, id2,   ...],  # Corresponding ProteinNet IDs
                  },
        "valid-10": {...},
            ...
        "valid-90": {...},
        "test":     {...},
            ...                              # Metadata
        }
\end{minted}

By default, the \texttt{load} function downloads the data from the web into the current directory and loads it as a Python dictionary. If the data already exists locally, it reads it from disk. Other than the requirement that the data must be loaded using Python, this method of data loading is agnostic to any downstream analyses.

\begin{minted}{python}
>>> import sidechainnet as scn
>>> data = scn.load(casp_version=12)
>>> data.keys()
['train', 'valid-10',...'valid-90', 'test', ...]
>>> data['train'].keys()
['seq', 'ang', 'ids', 'crd', 'msk', 'evo']
\end{minted}

\hypertarget{fourtwo}{}
\subsection{Loading SidechainNet with PyTorch DataLoaders}
The \texttt{load} function can also be used to load SidechainNet data as a dictionary of \texttt{torch.utils.data. DataLoader} objects. PyTorch \texttt{DataLoader}s make it simple to iterate over dataset items for training machine learning models. This method is recommended for using SidechainNet data with PyTorch. 

By default, the provided \texttt{DataLoader}s use a custom batching method that randomly generates batches of proteins of similar length. For efficient GPU usage, it generates larger batches when the average length of proteins in the batch is small and smaller batches when the proteins are large. The probability of selecting small-length batches is decreased so that each protein in SidechainNet is included in a batch with equal probability. See \texttt{dynamic\_batch\_size} and \texttt{collate\_fn} arguments for more information on modifying this behavior. In the example below, \texttt{model\_input} is a batched Tensor containing concatenated sequence and PSSM information.

\begin{minted}{python}

>>> dataloaders = scn.load(casp_version=12, with_pytorch="dataloaders")
>>> dataloaders.keys()
['train', 'train_eval', 'valid-10', ..., 'valid-90', 'test']
>>> dataloaders['train'].dataset
ProteinDataset(casp_version=12, split='train', n_proteins=81454,
               created='Sep 20, 2020')
>>> for protein_id, model_input, true_angles, true_coords in dataloaders['train']:
....    prediction = model(model_input)
....    loss = compute_loss(prediction, true_angles, true_coords)
....    ...

\end{minted}

We have also made it possible to access the protein sequence, mask, and PSSM data directly when training by adding \texttt{aggregate\_model\_input=False} to \texttt{scn.load}.

\begin{minted}{python}
>>> dataloaders = scn.load(casp_version=12, with_pytorch="dataloaders",
                           aggregate_model_input=False)
>>> for (protein_id, sequence, mask, pssm, true_angles,
         true_coords) in dataloaders['train']:
....    prediction = model(sequence, pssm)
....    ...
\end{minted}

\subsection{Converting Angle Representations into All-Atom Structures}
An important component of this work is the inclusion of both angular and 3D coordinate representations of each protein. Researchers who develop methods that rely on angular representations may be interested in converting this information into 3D coordinates. For this reason, SidechainNet provides a method to convert the angles it provides into Cartesian coordinates.

In the below example, \texttt{angles} is a NumPy matrix or Torch Tensor following the same organization as the NumPy angle matrices provided in SidechainNet. \texttt{sequence} is a string representing the protein's amino acid sequence. Both of these are obtainable from the SidechainNet data structures described in Sections \hyperlink{fourone}{4.1} and \hyperlink{fourtwo}{4.2}.

\begin{minted}{python}
>>> (len(sequence), angles.shape)  # 12 angles per residue
(128, (128, 12))
>>> sb = scn.StructureBuilder(sequence, angles)
>>> coords = sb.build()
>>> coords.shape  # 14 atoms per residue (128*14 = 1792)
(1792, 3)
\end{minted}

\subsection{Visualizing All-Atom Structures with PDB, py3Dmol, and gLTF Formats}
SidechainNet makes it easy to visualize both existing and predicted all-atom protein structures, as well as structures represented as either angles or coordinates.  These visualizations are available as PDB files, \texttt{py3Dmol.view} objects, and Graphics Library Transmission Format (gLTF) files. Examples of each are included below.

\begin{figure}
	\centering
    \includegraphics[width=1\linewidth]{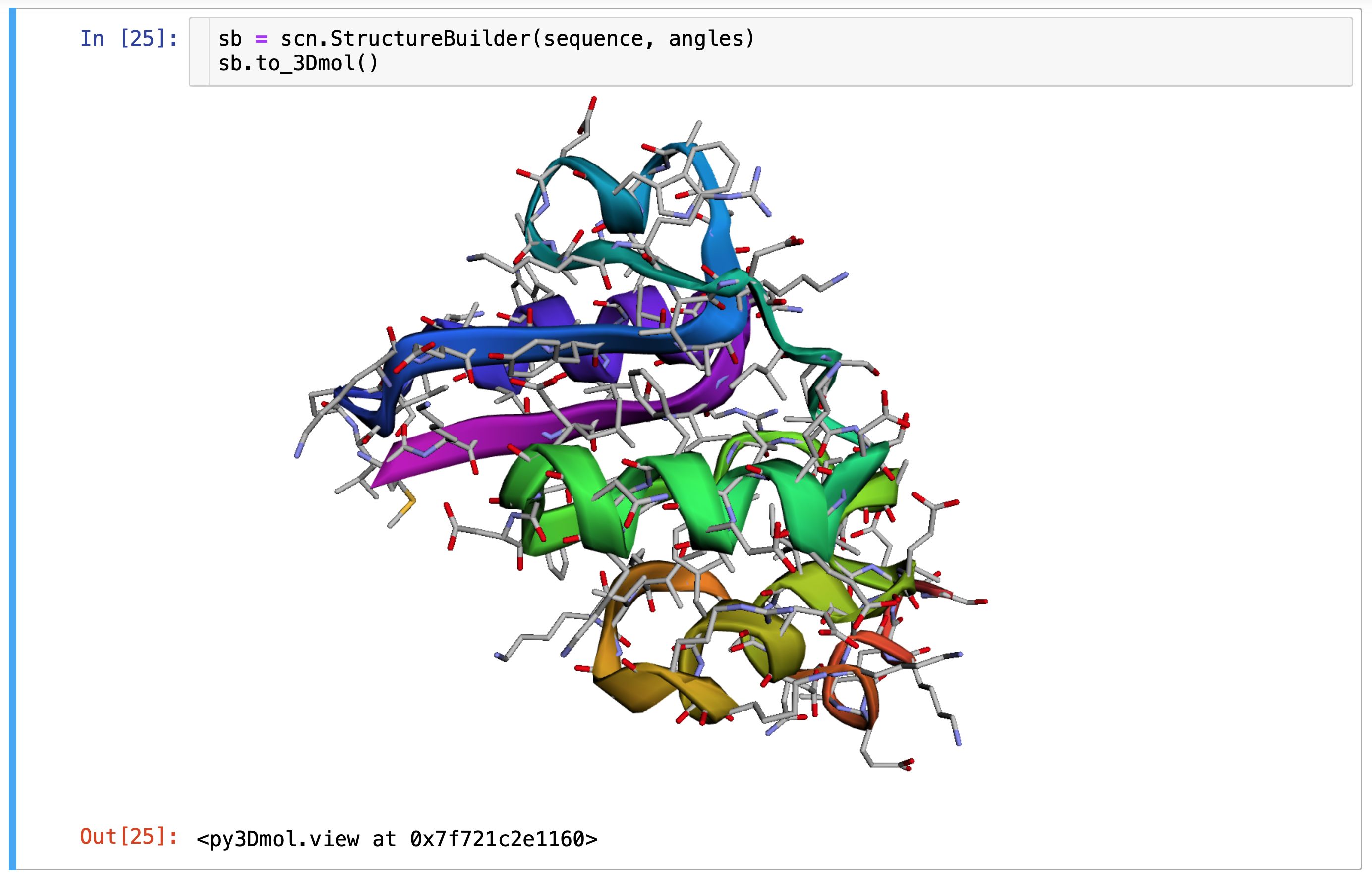}
	\caption{\texttt{py3Dmol} enables the user to interactively inspect all-atom protein structures from SidechainNet within Jupyter/iPython Notebooks.}
	\label{fig:3dmol}
\end{figure}

The PDB format is a typical format for representing protein structures and can be opened in software tools like PyMOL \cite{PyMOL}. \texttt{py3Dmol} (built on \href{http://3dmol.csb.pitt.edu}{3Dmol.js} \cite{Rego2014}) enables users to visualize and interact with protein structures on the web and in Jupyter Notebooks via an open-source, object-oriented, and hardware-accelerated Javascript library (Figure \ref{fig:3dmol}). Finally, gLTF files, despite their larger size, can be convenient for visualizing proteins on the web or in contexts where other protein visualization tools are not supported. 

\begin{minted}{python}
>>> sb.to_pdb("example.pdb")
>>> sb.to_gltf("example.gltf")
\end{minted}

\section{Discussion}
SidechainNet builds upon the strong foundation of ProteinNet to provide a protein structure dataset with minimal bias and maximum information content. By including many of the angles necessary to accurately reconstruct atomic Cartesian coordinates from an angle-only protein representation, we also enable future users to develop methods that utilize either or both internal and external coordinate systems at a previously inaccessible level of atomic detail. In addition, the tools developed for SidechainNet's programmatic construction may also be adapted for the generation of similar all-atom datasets with alternative methods of data clustering (e.g., datasets based on CATH \cite{Sillitoe2018}).

Methods developed to explicitly include the orientation and atomic coordinates of protein sidechains may have the upper hand in tasks such as structure-based drug discovery or in the analysis of specific enzyme activities, where such information is necessary. Although the impact of including sidechain information on existing predictive methods has yet to be studied, we hope SidechainNet makes such research more accessible.

SidechainNet code can be found at \texttt{\href{https://github.com/jonathanking/sidechainnet}{https://github.com/jonathanking/sidechainnet}}.

\begin{ack}
Many thanks to Mohammed AlQuraishi for his inspiring work on protein structure prediction. Thanks, also, to \href{https://github.com/JeppeHallgren}{Jeppe Hallgren} for his development of a ProteinNet text record parser, which I have used in part here.

 This work is supported by R01GM108340 from the National Institute of General Medical Sciences, is supported in part by the University of Pittsburgh Center for Research Computing through the resources provided, and by NIH T32 training grant T32 EB009403 as part of the HHMI-NIBIB Interfaces Initiative.
\end{ack} 
 
\small
\bibliographystyle{unsrt}
\bibliography{main}

\begin{thebibliography}{10}

\bibitem{AlphaFoldNews}
Matthew Hutson.
\newblock {AI} protein-folding algorithms solve structures faster than ever.
\newblock {\em Nature}, July 2019.

\bibitem{Senior2020}
Andrew~W. Senior, Richard Evans, John Jumper, James Kirkpatrick, Laurent Sifre,
  Tim Green, Chongli Qin, Augustin {\v{Z}}{\'{\i}}dek, Alexander W.~R. Nelson,
  Alex Bridgland, Hugo Penedones, Stig Petersen, Karen Simonyan, Steve Crossan,
  Pushmeet Kohli, David~T. Jones, David Silver, Koray Kavukcuoglu, and Demis
  Hassabis.
\newblock Improved protein structure prediction using potentials from deep
  learning.
\newblock {\em Nature}, 577(7792):706--710, January 2020.

\bibitem{CASP13}
Andriy Kryshtafovych, Torsten Schwede, Maya Topf, Krzysztof Fidelis, and John
  Moult.
\newblock Critical assessment of methods of protein structure prediction
  ({CASP}){\textemdash}round {XIII}.
\newblock {\em Proteins: Structure, Function, and Bioinformatics},
  87(12):1011--1020, October 2019.

\bibitem{ProteinRep}
Ethan~C. Alley, Grigory Khimulya, Surojit Biswas, Mohammed AlQuraishi, and
  George~M. Church.
\newblock Unified rational protein engineering with sequence-based deep
  representation learning.
\newblock {\em Nature Methods}, 16(12):1315--1322, October 2019.

\bibitem{ingraham2019generative}
John Ingraham, Vikas~K Garg, Regina Barzilay, and Tommi Jaakkola.
\newblock Generative models for graph-based protein design.
\newblock In {\em Advances in Neural Information Processing Systems}, 2019.

\bibitem{SureyyaRifaioglu2019}
Ahmet~Sureyya Rifaioglu, Tunca Do{\u{g}}an, Maria~Jesus Martin, Rengul
  Cetin-Atalay, and Volkan Atalay.
\newblock {DEEPred}: Automated protein function prediction with multi-task
  feed-forward deep neural networks.
\newblock {\em Scientific Reports}, 9(1), May 2019.

\bibitem{DeepFam}
Seokjun Seo, Minsik Oh, Youngjune Park, and Sun Kim.
\newblock {DeepFam}: deep learning based alignment-free method for protein
  family modeling and prediction.
\newblock {\em Bioinformatics}, 34(13):i254--i262, June 2018.

\bibitem{Xu2019}
Jinbo Xu.
\newblock Distance-based protein folding powered by deep learning.
\newblock {\em Proceedings of the National Academy of Sciences},
  116(34):16856--16865, August 2019.

\bibitem{Yang2020}
Jianyi Yang, Ivan Anishchenko, Hahnbeom Park, Zhenling Peng, Sergey
  Ovchinnikov, and David Baker.
\newblock Improved protein structure prediction using predicted interresidue
  orientations.
\newblock {\em Proceedings of the National Academy of Sciences},
  117(3):1496--1503, January 2020.

\bibitem{AlQuraishi2019}
Mohammed AlQuraishi.
\newblock End-to-end differentiable learning of protein structure.
\newblock {\em Cell Systems}, 8(4):292--301.e3, April 2019.

\bibitem{ingraham2018learning}
John Ingraham, Adam Riesselman, Chris Sander, and Debora Marks.
\newblock Learning protein structure with a differentiable simulator.
\newblock In {\em International Conference on Learning Representations}, 2019.

\bibitem{imagenet}
Olga Russakovsky, Jia Deng, Hao Su, Jonathan Krause, Sanjeev Satheesh, Sean Ma,
  Zhiheng Huang, Andrej Karpathy, Aditya Khosla, Michael Bernstein,
  Alexander~C. Berg, and Li~Fei-Fei.
\newblock {ImageNet Large Scale Visual Recognition Challenge}.
\newblock {\em International Journal of Computer Vision (IJCV)},
  115(3):211--252, 2015.

\bibitem{uniprot}
{The UniProt Consortium}.
\newblock {UniProt}: a worldwide hub of protein knowledge.
\newblock {\em Nucleic Acids Research}, 47(D1):D506--D515, November 2018.

\bibitem{ThePDB}
H.~M. Berman.
\newblock The protein data bank.
\newblock {\em Nucleic Acids Research}, 28(1):235--242, January 2000.

\bibitem{standards}
David~T. Jones.
\newblock Setting the standards for machine learning in biology.
\newblock {\em Nature Reviews Molecular Cell Biology}, 20(11):659--660,
  September 2019.

\bibitem{dude}
Michael~M. Mysinger, Michael Carchia, John.~J. Irwin, and Brian~K. Shoichet.
\newblock Directory of useful decoys, enhanced ({DUD}-e): Better ligands and
  decoys for better benchmarking.
\newblock {\em Journal of Medicinal Chemistry}, 55(14):6582--6594, July 2012.

\bibitem{dudeissues}
Lieyang Chen, Anthony Cruz, Steven Ramsey, Callum~J. Dickson, Jose~S. Duca,
  Viktor Hornak, David~R. Koes, and Tom Kurtzman.
\newblock Hidden bias in the {DUD}-e dataset leads to misleading performance of
  deep learning in structure-based virtual screening.
\newblock {\em {PLOS} {ONE}}, 14(8):e0220113, August 2019.

\bibitem{sieg2019need}
Jochen Sieg, Florian Flachsenberg, and Matthias Rarey.
\newblock In need of bias control: evaluating chemical data for machine
  learning in structure-based virtual screening.
\newblock {\em Journal of chemical information and modeling}, 59(3):947--961,
  2019.

\bibitem{scpacking}
Md~Shariful~Islam Bhuyan and Xin Gao.
\newblock A protein-dependent side-chain rotamer library.
\newblock {\em {BMC} Bioinformatics}, 12(S14), December 2011.

\bibitem{rosettaCASP8}
Srivatsan Raman, Robert Vernon, James Thompson, Michael Tyka, Ruslan Sadreyev,
  Jimin Pei, David Kim, Elizabeth Kellogg, Frank DiMaio, Oliver Lange, Lisa
  Kinch, Will Sheffler, Bong-Hyun Kim, Rhiju Das, Nick~V. Grishin, and David
  Baker.
\newblock Structure prediction for {CASP}8 with all-atom refinement using
  rosetta.
\newblock {\em Proteins: Structure, Function, and Bioinformatics},
  77(S9):89--99, 2009.

\bibitem{Shapovalov2011}
Maxim~V. Shapovalov and Roland~L. Dunbrack.
\newblock A smoothed backbone-dependent rotamer library for proteins derived
  from adaptive kernel density estimates and regressions.
\newblock {\em Structure}, 19(6):844--858, June 2011.

\bibitem{proteinnet}
Mohammed AlQuraishi.
\newblock {ProteinNet}: a standardized data set for machine learning of protein
  structure.
\newblock {\em {BMC} Bioinformatics}, 20(1), June 2019.

\bibitem{amber}
David~A. Case, Thomas~E. Cheatham, Tom Darden, Holger Gohlke, Ray Luo,
  Kenneth~M. Merz, Alexey Onufriev, Carlos Simmerling, Bing Wang, and Robert~J.
  Woods.
\newblock The amber biomolecular simulation programs.
\newblock {\em Journal of Computational Chemistry}, 26(16):1668--1688, 2005.

\bibitem{amberff19sb}
Chuan Tian, Koushik Kasavajhala, Kellon A.~A. Belfon, Lauren Raguette,
  He~Huang, Angela~N. Migues, John Bickel, Yuzhang Wang, Jorge Pincay, Qin Wu,
  and Carlos Simmerling.
\newblock ff19sb: Amino-acid-specific protein backbone parameters trained
  against quantum mechanics energy surfaces in solution.
\newblock {\em Journal of Chemical Theory and Computation}, 16(1):528--552,
  November 2019.

\bibitem{Bakan2011}
A.~Bakan, L.~M. Meireles, and I.~Bahar.
\newblock {ProDy}: Protein dynamics inferred from theory and experiments.
\newblock {\em Bioinformatics}, 27(11):1575--1577, April 2011.

\bibitem{PyMOL}
{Schr\"odinger, LLC}.
\newblock The {PyMOL} molecular graphics system, version~1.8.
\newblock November 2015.

\bibitem{Rego2014}
N.~Rego and D.~Koes.
\newblock 3dmol.js: molecular visualization with {WebGL}.
\newblock {\em Bioinformatics}, 31(8):1322--1324, December 2014.

\bibitem{Sillitoe2018}
Ian Sillitoe, Natalie Dawson, Tony~E Lewis, Sayoni Das, Jonathan~G Lees, Paul
  Ashford, Adeyelu Tolulope, Harry~M Scholes, Ilya Senatorov, Andra Bujan,
  Fatima~Ceballos Rodriguez-Conde, Benjamin Dowling, Janet Thornton, and
  Christine~A Orengo.
\newblock {CATH}: expanding the horizons of structure-based functional
  annotations for genome sequences.
\newblock {\em Nucleic Acids Research}, 47(D1):D280--D284, November 2018.

\end{thebibliography}

\end{document}